\documentstyle[preprint,aps]{revtex}

\newcommand{\be}{\begin{equation}}
\newcommand{\ee}{\end{equation}}
\newcommand{\bea}{\begin{eqnarray}}
\newcommand{\eea}{\end{eqnarray}}
\newcommand{\bem}{\begin{mathletters}}
\newcommand{\eem}{\end{mathletters}}

\newcommand{\sla}{\! \not \!}
\newcommand{\otto}{\leftrightarrow}
\newcommand{\Punkt}{\quad .}
\newcommand{\Komma}{\quad ,}

\begin{document}

\preprint{   }

\title{Strangeness Production Incorporating Chiral Symmetry
\thanks{To appear in the proceedings of the 4th International Symposium on
Strangeness in Quark Matter, Padova, Italy, July 20--24.} }

\author{P.~Rehberg}
\address{SUBATECH \\
         Laboratoire de Physique Subatomique et des Technologies
                  Associ\'ees \\
         UMR Universt\'e de Nantes, IN2P3/CNRS, Ecole des Mines de Nantes \\
         4 Rue Alfred Kastler, F-44070 Nantes Cedex 3, France}

\maketitle
\vspace{4cm}
\begin{abstract}
Chiral symmetry is known to be decisive for an understanding of the
low energy sector of strong interactions. It is thus important for a
model of relativistic heavy ion collisions to incorporate the dynamical
breaking and restoration of chiral symmetry. Thus we study an expansion
scenario for a quark-meson plasma using the Nambu--Jona-Lasinio (NJL)
model in its three flavor version. The equations of motion for light and
strange quarks as well as for pions, kaons and etas are solved using a
QMD type algorithm, which is based on a parametrization of the Wigner
function. The scattering processes incorporated into this calculation are
of the types $qq\otto qq$, $q\bar q \otto q\bar q$, $q\bar q\otto MM$
and $M\otto q\bar q$.
\end{abstract}

\clearpage

In relativistic heavy ion collisions, a very hot and dense system of
strongly interacting matter is created.  At these high temperatures
and densities, the state of matter changes drastically. The two main
effects which are predicted from quantum chromodynamics (QCD) are (i)
deconfinement and (ii) chiral symmetry restauration. Lattice simulations
indicate, that these two effects are not independent of each other, but
occur at the same time \cite{edwin}.  Although experiments searching for
this new state of matter have been performed at the Cern SPS and will be
performed at RHIC and LHC, the theoretical description still leaves many
open questions. Some of these are (i) the question, if a local thermal
and/or chemical equilibrium will be established, (ii) the influence of the
in-medium changes of hadronic properties and (iii) the description of the
hadronization.  The treatment of these problems is by no means simple,
since a satisfactory, simple model for the description of the phase
transition to a quark-gluon plasma (QGP) does not yet exist. Since the
QGP, if created, is a transient state, a realistic model for its evolution
should be able to handle nonequilibrium effects. A nonequilibrium theory
of nonabelian gauge theories such as QCD, however, cannot be provided
using present days knowledge, since these theories even in thermal
equilibrium are not yet sufficiently understood. Lattice gauge theories,
on the other hand, have been successfully applied in order to extract
the thermal behaviour of strongly interacting matter. They work,
however, only in equilibrium situations. Another practical approach
which is frequently used is hydrodynamics. These models, however,
suffer from the fact that they work in the limit of infinitely
many collisions per time interval, which might not be true in the
first stages and surely is not true in the late stages of the collision.

The ansatz presented here thus the modelization of the expansion of a
hot system using an effective Lagrangian. The model interaction we use
is the three flavor Nambu--Jona-Lasinio (NJL) model \cite{sandi},
defined by
\bea
{\cal L} &=& \sum_{f=u,d,s} \bar\psi_f (i\sla\partial - m_{0f}) \psi_f
+G \sum_{a=0}^8 \left[\left(\bar\psi\lambda_a\psi\right)^2
+\left(\bar\psi i\gamma_5\lambda_a\psi\right)^2 \right]
\label{lagra} \\ \nonumber
&+& K\left[ \det \bar\psi\left(1+\gamma_5\right)\psi +
\det \bar\psi\left(1-\gamma_5\right)\psi \right]
\Punkt
\eea
In this model, the exchange interaction has been contracted to give a
pointlike interaction in flavor space. This approximation works as long
as the effective mass of the gluon is high. Recent QCD calculations
support this assumption even at high temperatures \cite{kajantie}.
The most important feature of the Lagrangian (\ref{lagra}) is that it
preserves the chiral symmetry of QCD, i.\,e. in the limit
$m_{0f}\to 0$ it is invariant under transformations of the form
\be
\psi \to \exp\left(-i\theta_a\lambda_a\gamma_5\right) \psi
\Punkt
\ee
It is a well established fact, that this symmetry is the essential
ingredient for the description of the low lying hadronic states
which are produced copiously in a heavy ion collision \cite{dono}.

The equilibrium properties of the NJL model in thermal
equilibrium have been studied in great detail elsewhere
\cite{sandi,su2thermo,gerry,su3elast,su3hadron}, so that only those
properties will be mentioned here, which are important for the
understanding of the present article. For simplicity, first the case
$m_{0f}=0$ will be detailled. At temperature $T=0$, chiral symmetry is
spontaneously broken, which leads to a finite effective quark mass. As a
consequence of the Goldstone theorem, $N_f^2-1$ massless modes appear as
quark-antiquark bound states. These states are identified with the $\pi$,
$K$ and $\eta$ mesons. At a certain finite temperature, however, chiral
symmetry gets restored and one has again massless quarks, whereas the
mesons become massive. In this phase, they do no longer exist as bound
states but rather as resonances. If one introduces a finite current quark
mass, $m_{0f}\ne 0$, this picture gets slightly changed in that chiral
symmetry is no longer an exact, but rather an approximate symmetry. Thus
mesons have a small but finite mass at $T=0$, as they have in nature,
whereas the quark mass does not go to zero at large temperatures, but
stays finite and for $T\to\infty$ goes to the current quark mass.

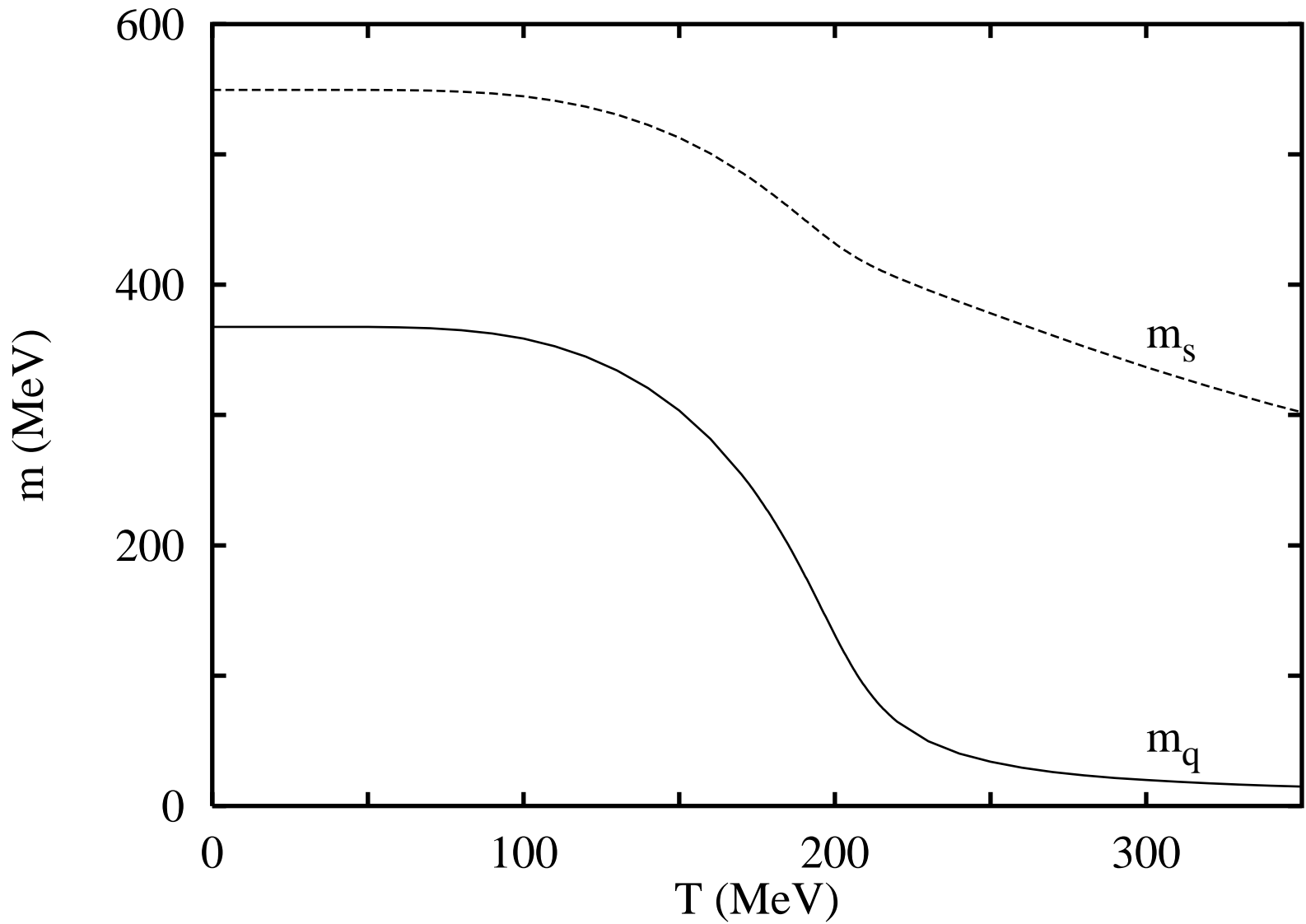
\begin{figure}
\caption[]{Quark masses at finite temperature in equilibrium. Solid line:
           light quarks, dashed line: strange quarks.}
\label{qmass}
\end{figure}

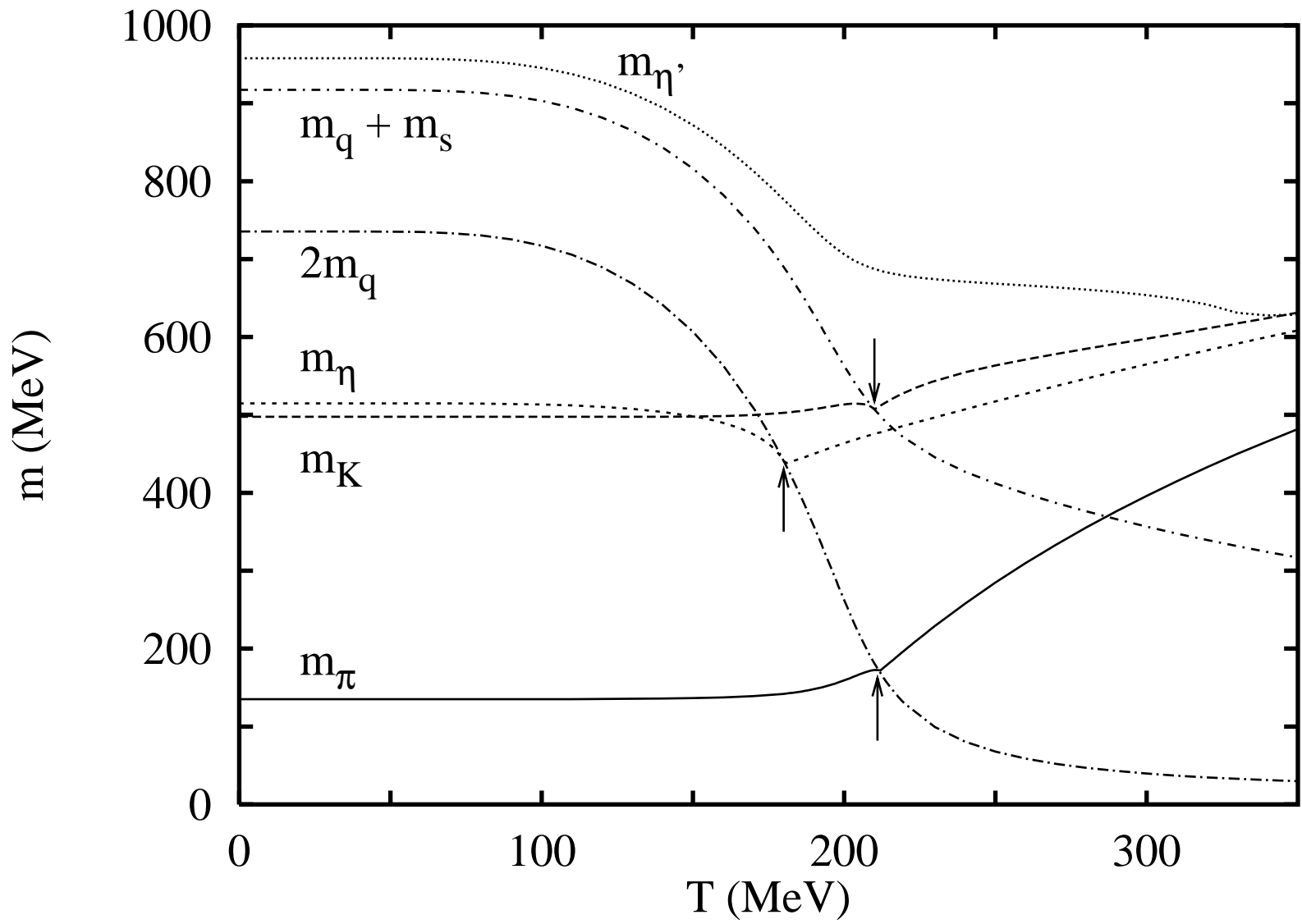
\begin{figure}
\caption[]{Meson masses as a function of temperature. Also shown are
           $2m_q$ and $m_q+m_s$. The Mott transitions of the pion, the
           kaon and the eta are marked by the arrows.}
\label{mmass}
\end{figure}

The quark mass spectrum for this case is shown in Fig.~\ref{qmass}.
The light quarks $u$ and $d$ are taken to be degenerate and will
be denoted by the generic index $q$ in the following. The current
quark masses used in Fig.~\ref{qmass} are $m_{0q}=5.5$\,MeV and
$m_{0s}=140$\,MeV. Due to chiral symmetry breaking, one has at $T=0$ an
effective mass of $m_q=368$\,MeV and $m_s=550$\,MeV.  In the temperature
region $T\approx 200$\,MeV, these masses drop and the effective mass
of the light quarks becomes low. For the strange quarks, however,
the situation is different. Due to the high current quark mass, the
effective mass stays relatively high.  At $T=350$\,MeV, which is far
beyond any temperature to be realistically expected in present days heavy
ion experiments, one still has $m_s=300$\,MeV, which is about twice
as high as the current quark mass. The pseudoscalar mesonic mass spectrum
using the same parameters is shown in Fig.~\ref{mmass}. At $T=0$, the
mesons have their physical masses of $m_\pi=135$\,MeV, $m_K=498$\,MeV
and $m_\eta=515$\,MeV.  With rising temperatures, these values stay more
or less constant, until to a certain temperature, where their masses
become equal to the masses of their constituents, i.\,e. $m_\pi=2m_q$,
$m_K=m_q+m_s$ or $m_\eta=2m_q$, respectively. This happens when the
$m_\pi$ and $m_\eta$ lines of Fig.~\ref{mmass} cross the $2m_q$ line, or
the $m_K$ line crosses the $m_q+m_s$ line, respectively, as is indicated
by the arrows.  At these temperatures, the mesons become unstable via a
Mott transition \cite{gerry}. At higher temperatures, they are unstable
with respect to the decay $M\to q\bar q$ and obtain a finite width.
The qualitative form of the mass spectrum of Fig.~\ref{mmass} has been
confirmed by lattice calculations \cite{edwin}.

A finite temperature study of this model will surely not lead to an event
generator, since the model still contains free quark states, which are
not observed in nature. It will, however, be able to study qualitative
features of the plasma expansion, such as expansion time scales, the
approach to equilibrium and thus the applicability of hydrodynamics,
the production mechanism of hadrons etc. While the simulation program
presently is limited to zero baryochemical potential, future extensions
will remove this limitation and also allow for the study of strangeness
destillation and DCC formation.

The treatment of the NJL model in nonequilibrium starts from the
observation, that both quark and meson degrees of freedom can
be {\em simultaneously \/} described by an equation of the
Boltzmann type \cite{pion},
\be
\left(\partial_t + \vec\partial_pE\vec\partial_x
                 - \vec\partial_xE\vec\partial_p\right) n(t, \vec x, \vec p)
= I_{\rm coll}\left[n(t, \vec x, \vec p)\right]
\label{boltz}\Komma
\ee
where $I_{\rm coll}[n]$ is a collision integral. In principle, one
is thus able to describe a transition from a pure quark regime to a
hadronic regime, where quarks are converted to hadrons via collision
processes. The solution method we choose for Eq.~(\ref{boltz}) is an
algorithm of the QMD type \cite{qmd}, i.\,e. we parametrize the particle
distribution functions by
\be
n(t, \vec x, \vec p)
   = \sum_i \exp\left(-\frac{(\vec x - \vec x_i(t))^2}{2w^2}\right)
            \exp\left(-\frac{w^2}{2}(\vec p - \vec p_i(t))^2\right)
\Punkt
\ee
The center points of the distribution move on the characteristics of
Eq.~(\ref{boltz}), i.\,e.
\be
\dot{\vec{x}}_i (t) = \vec p_i(t) / E \hspace{2cm}
\dot{\vec{p}}_i (t) = - \vec \partial_x E + \mbox{collision contributions}
\label{move} \Punkt
\ee
Equation (\ref{move}) has  to be solved together with the gap equation
\bem \label{gap} \bea
m_i &=& m_{0i} - \frac{GN_c}{\pi^2} m_i A_i 
         + \frac{KN_c^2}{8\pi^4} m_j A_j m_k A_k, \qquad i\ne j\ne k \ne i \\
A_i &=& -8 \pi^2 \int \frac{d^3p}{(2\pi)^3} \frac{1}{E}
        \left(1- \frac{n_i + n_{\bar i}}{2N_c}\right) \Komma
\eea \eem
where the indices $i$, $j$ and $k$ run over all three quark flavors. For
the computation of mesonic properties we take a shortcut by defining
an effective temperature via $m_q(\vec x, t) = m_q^{\rm eq}(T_{\rm
eff}(\vec x, t))$, where $m_q^{\rm eq}(T)$ is the equilibrium form of the
temperature dependence of $m_q$ as shown in Fig.~\ref{qmass}.  The mesonic
properties can then be obtained using the equilibrium expressions, which
are functions of $T_{\rm eff}$, $m_q$ and $m_s$. The collision processes
entering Eq.~(\ref{move}) are (i) quark quark scattering $qq\otto qq$,
$q\bar q\otto q\bar q$ and $\bar q\bar q\otto\bar q\bar q$
\cite{su3elast}, (ii) hadronization $q\bar q\otto MM$ \cite{su3hadron},
and (iii) meson decay $M\to q\bar q$ \cite{pion}. The latter process is
only possible in the early phases of the expansion, when the effective
temperature is higher than the Mott temperature.

The initial conditions chosen presently do not contain strange quarks,
whereas light quarks are distributed thermally within a sphere of a given
radius. This kind of initial conditions has the immediate consequence,
that the strange quark mass is {\em even higher \/} than in thermal
equilibrium.  The reason for this can be seen by writing Eq.~(\ref{gap})
explicitly for strange quarks. One recognizes that the second term on
the right hand side, which is proportional to $G$, only couples to the
strange quark condensate. Since initially no strange quarks are present,
this term does not receive medium corrections and thus makes the effective
quark mass higher than in thermal equilibrium. Medium corrections do
arise from the third term of Eq.~(\ref{gap}), which is proportional
to the product of the up and down quark condensate. This contribution
is, however, not sufficient to produce a large mass drop.

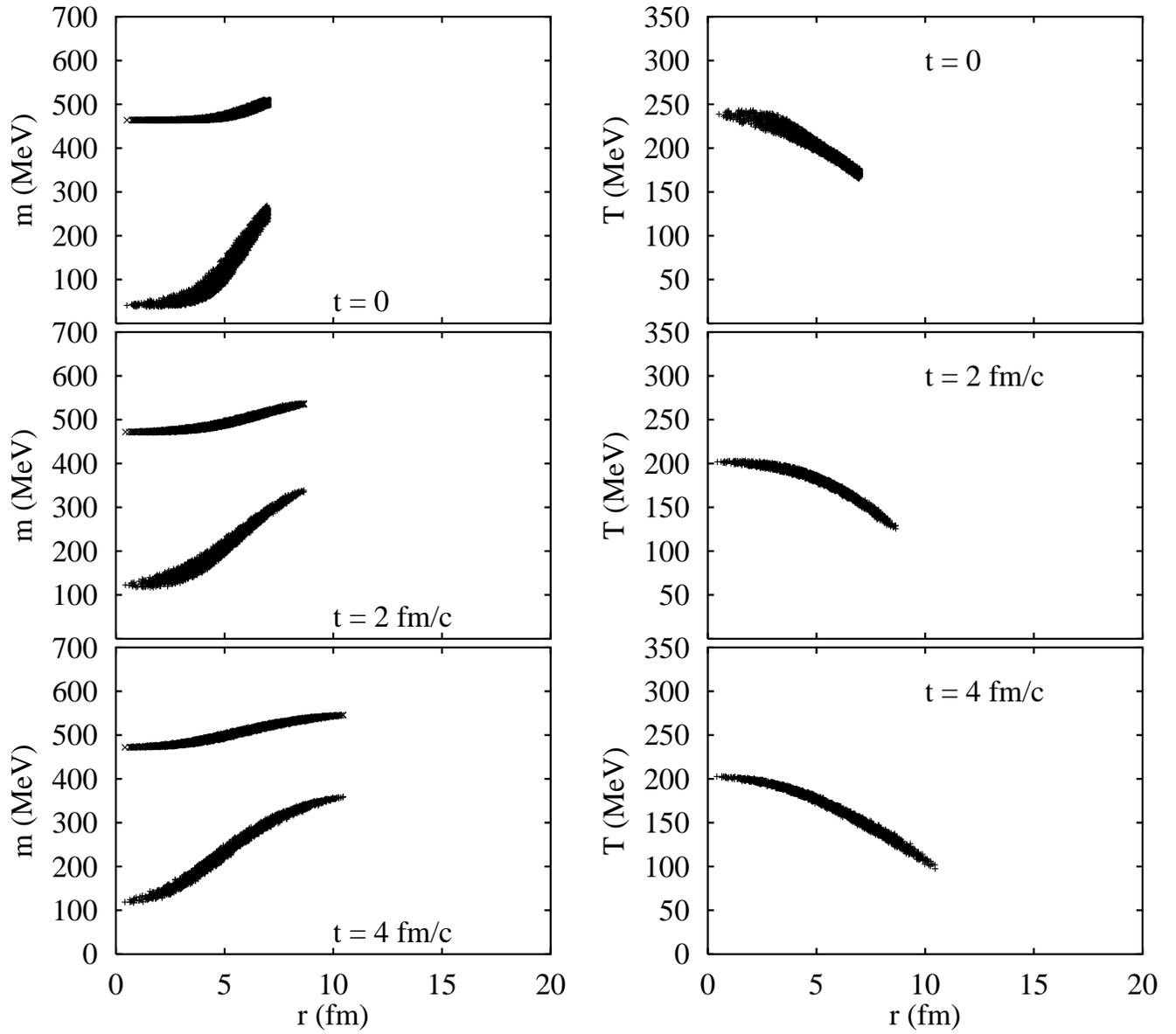
\begin{figure}
\caption[]{Effective quark masses (left) and temperatures (right) as a
           function of $r$ for $t=0$, 2 and 4\,fm$/c$. In the left panel,
           the upper curves are for strange quarks and the lower ones
           for light quarks, respectively.}
\label{massev}
\end{figure}

Quantitatively, this can be seen from Fig.~\ref{massev}, which shows
the quark masses and the effective temperatures as a function of $r$
at various times during the initial phase of the expansion. At $t=0$,
the light quark mass in the center of the system is low, according to
the high particle density here. At larger radii, the mass goes up due
to the gaussian form of the particle distribution. Note that the quark
masses are only known at those points where a particle is present, thus
the curves stop at the edge of the fireball. The effective temperature
in the centre amounts to approximately 250\,MeV. At this temperature in
equilibrium, one would expect a strange quark mass of 380\,MeV according
to Fig.~\ref{qmass}.  In reality, however, one has a strange quark mass
of approximately 450\,MeV in the centre due to nonequilibrium effects. At
later times, the system expands and thus the quark mass curve becomes
flatter, while its width grows. Accordingly, the effective temperature
drops.  In the final state, the quark masses will tend towards the vacuum
value, which means that the mean field part of the interaction
ceases \cite{trap}.

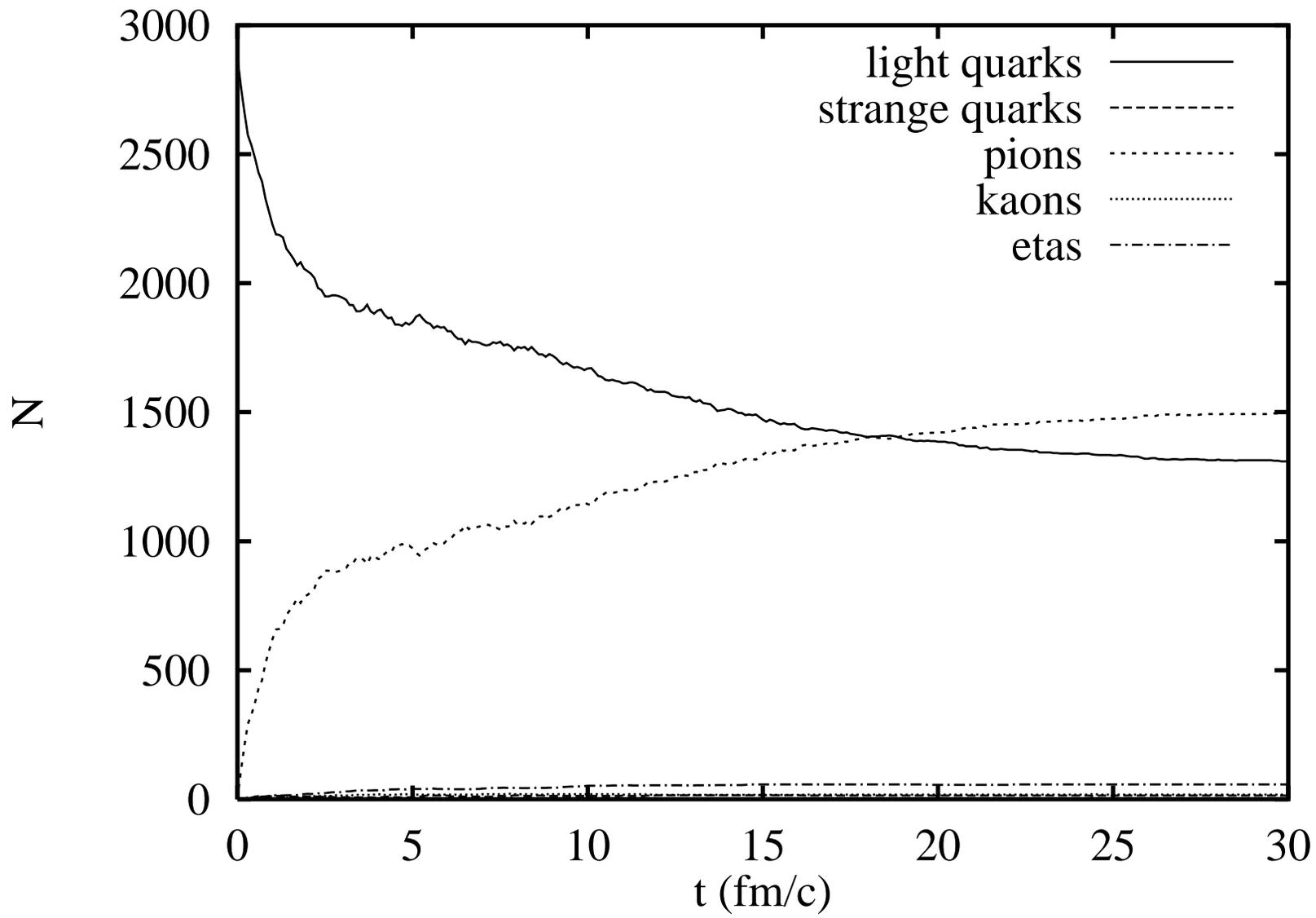
\begin{figure}
\caption[]{Time dependence of the particle multiplicities.}
\label{multi}
\end{figure}

Mesons are created during the evolution by collisions of quarks and
antiquarks. Most of all mesons created are pions, which are most
frequently produced by the collision of two light quarks. Kaons,
on the other hand, are produced most easily by the collision of
a light with a strange quark or two strange quarks, while the
hadronization cross section for the creation of a kaon pair from a light
quark antiquark pair is rather low \cite{su3hadron}. Eta mesons
are most frequently created collisions of two light quarks, forming
a pion and an eta. The time dependence of the multiplicities is
shown in Fig.~\ref{multi}. It can be seen here, that the production of
mesons starts immediately after the beginning of the expansion.
The production rate is maximal at $t=0$, when the particle density
is high, giving thus rise to a large number of collisions. At later
times the density drops and thus the number of collisions per
time decreases. This in turn leads to a flattening of the multiplicity
curves. 

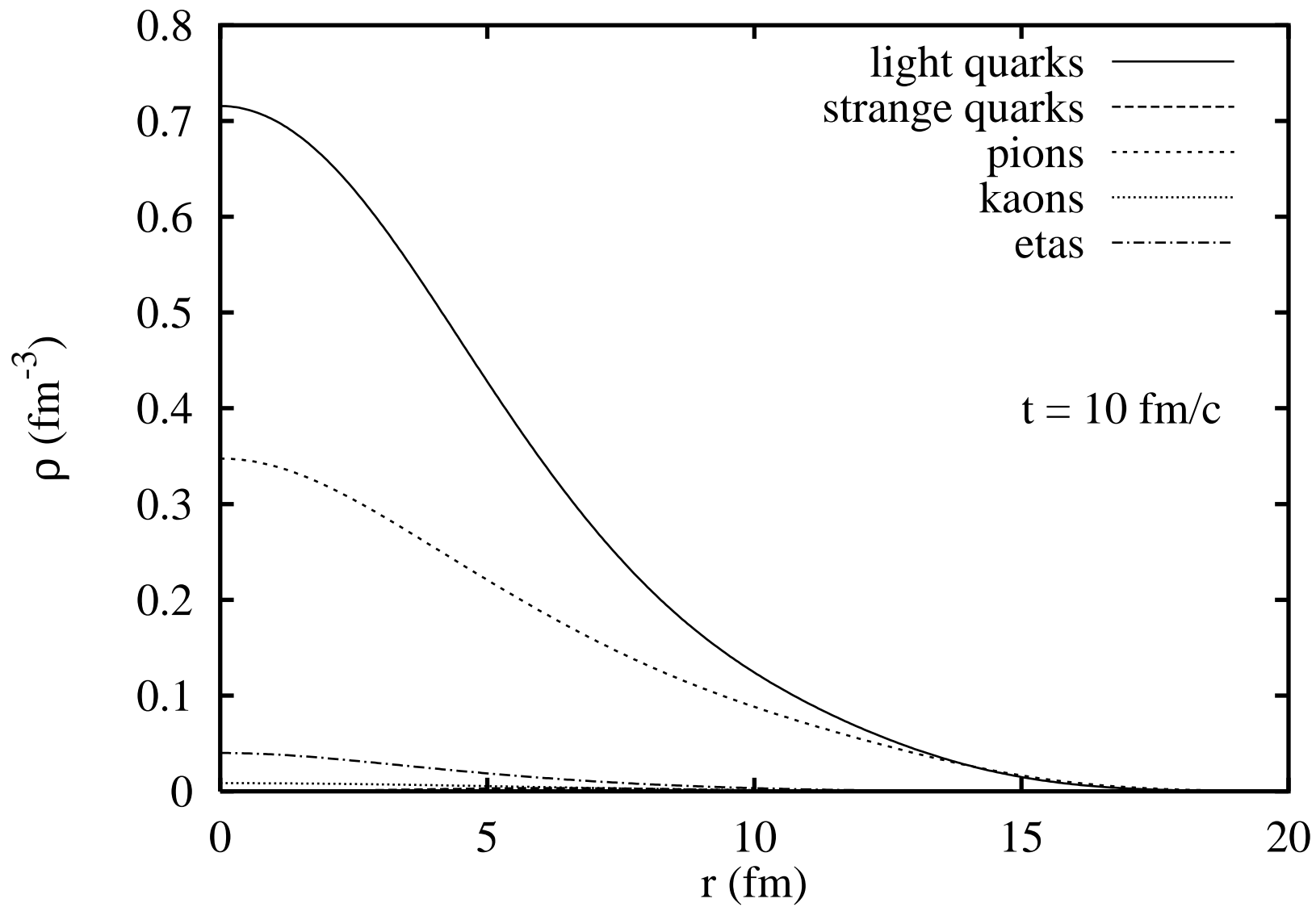
\begin{figure}
\caption[]{Angular averaged particle density at $t=10$\,fm$/c$.}
\label{densev}
\end{figure}

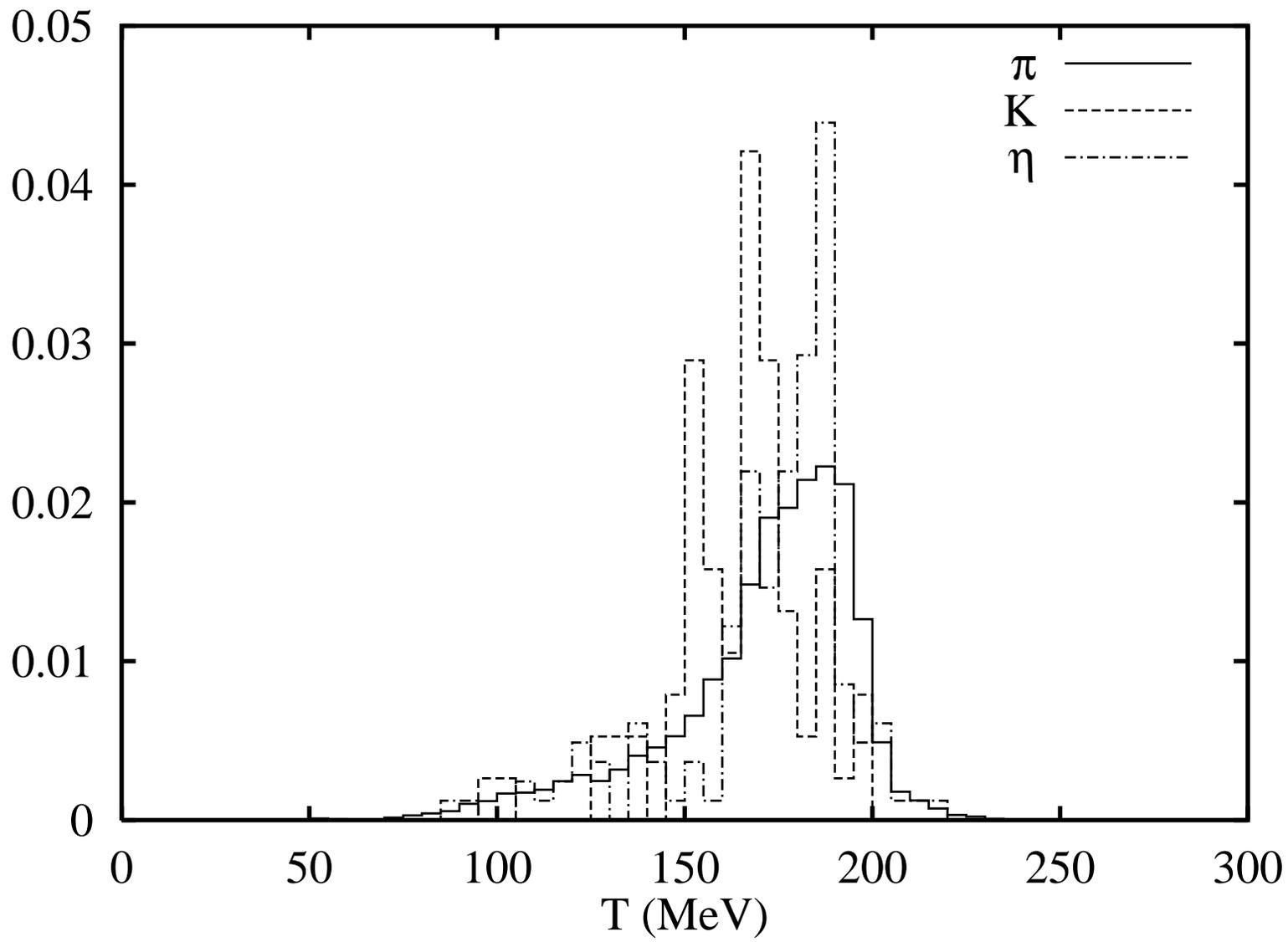
\begin{figure}
\caption[]{Histogram of the creation temperatures of pions (solid) line,
           kaons (dashed line) and etas (dot-dashed line).}
\label{thist}
\end{figure}

Figure \ref{densev} shows the particle density, averaged over all solid angles,
at time $t=10$\,fm$/c$. It is clearly visible that the meson density
is maximal at the same places where also the quark density is high.
This confirms the picture gained from Fig.~\ref{multi}, i.\,e. that
mesons are created within the bulk of the fireball. More insight
can be gained from Fig.~\ref{thist}, which shows the distribution of
the effective temperatures at the creation points of the mesons
for pions, kaons and etas respectively. All three distributions
agree more or less with each other, up to the effects of lower
statistics for the kaon and the eta. It can be seen that mesons are
created most likely at temperatures directly below the Mott temperature,
and within a temperature range of 50\,MeV. This plot should be compared
to the mean hadronization times for light and strange quarks in
equilibrium, as have been calculated in \cite{su3hadron}. It has
been shown there, that these hadronization times have a minimum
in the temperature range 150\,MeV$<T<$200\,MeV, which agrees nicely
with Fig.~\ref{thist}.

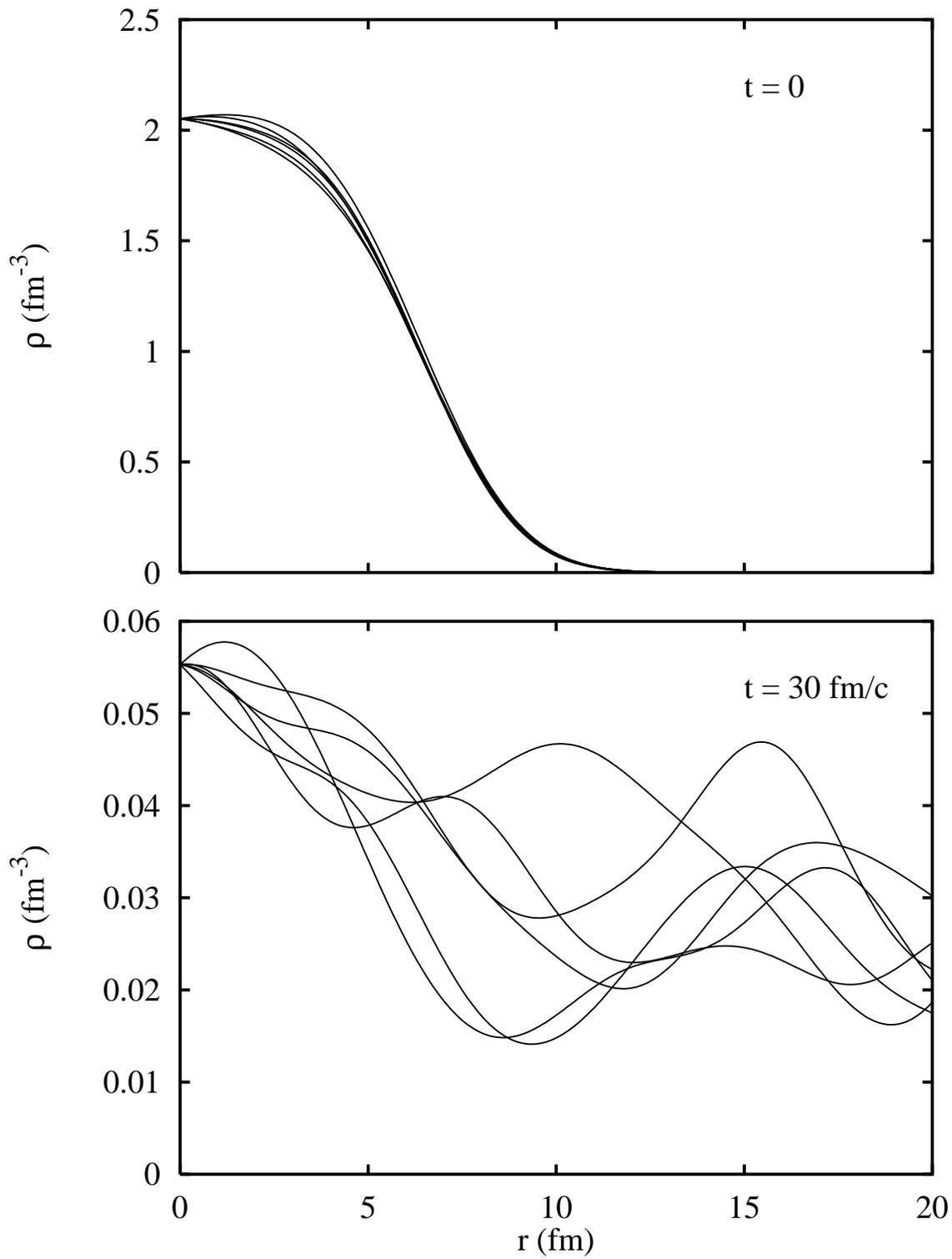
\begin{figure}
\caption[]{Particle density along the coordinate axes. Upper panel:
           $t=0$, lower panel: $t=30$\,fm$/c$.}
\label{dflukt}
\end{figure}

Figure \ref{dflukt} shows the density along the coordinate axes for
$t=0$ and $t=30$\,fm$/c$. It can be seen here, that, although one starts
with a system, which is to a good approximation spherically symmetric,
one ends up with a final state which shows large fluctuations of the
density with respect to the direction. This behaviour contradicts
hydrodynamics, where an initially symmetric system stays symmetric.

To conclude, it hat been demonstrated how a chirally symmetric
quark-meson plasma containing strangeness behaves out of equilibrium
and how mesons are produced. This investigation has been performed
at zero baryochemical potential. Future investigations will remove
this limitation and thus be able to study the mechanism of strangeness
destillation. Also the study of DCC formation is planned.

{\bf Acknowledgments.} Fruitful discussions with L.~Bot and J.~Aichelin
are greatfully acknowledged.

\begin{references}
\bibitem{edwin}
        E.~Laermann, Nucl. Phys. A {\bf 610}, 1c (1996);
        Nucl. Phys. B (Proc. Suppl.) {\bf 60A}, 180 (1998).
\bibitem{sandi}
        S.\,P.~Klevansky, Rev. Mod. Phys. {\bf 64}, 649 (1992);
        T.~Hatsuda and T.~Kunihiro, Phys. Rep. {\bf 247}, 221 (1994).
\bibitem{kajantie}
        K.~Kajantie, M.~Laine, J.~Peisa, A.~Rajantie, K.~Rummukainen
        and M.~Shaposhnikov, Phys. Rev. Lett. {\bf 79}, 3130 (1997);
        Nucl. Phys. (Proc. Suppl.) {\bf 63}, 418 (1998).
\bibitem{dono}
        J.\,F.~Donoghue, E.~Golowich and B.\,R.~Holstein, {\it Dynamics
        of the Standard Model} (Cambridge University Press, 1992).
\bibitem{su2thermo}
        J.~H\"ufner, S.\,P.~Klevansky, P.~Zhuang and H.~Voss,
        Ann. Phys. (NY) {\bf 234}, 225 (1993);
        P.~Zhuang, J.~H\"ufner and S.\,P.~Klevansky, Nucl. Phys. A
        {\bf 576}, 525 (1994).
\bibitem{gerry}
        J.~H\"ufner, S.\,P.~Klevansky and P.~Rehberg, Nucl. Phys. A {\bf
        606}, 260 (1996).
\bibitem{su3elast}
        P.~Rehberg, S.\,P.~Klevansky and J.~H\"ufner, Nucl. Phys. A {\bf
        608}, 356 (1996).
\bibitem{su3hadron}
        P.~Rehberg, S.\,P.~Klevansky and J.~H\"ufner, Phys. Rev. C {\bf
        53}, 410 (1996).
\bibitem{pion}
        P.~Rehberg, Phys. Rev. C {\bf 57}, 3299 (1998).
\bibitem{qmd}
        J.~Aichelin, Phys. Rep. {\bf 202}, 233 (1991).
\bibitem{trap}
        P.~Rehberg and J.~H\"ufner, Nucl. Phys. A {\bf 635}, 511 (1998).
\end{references}
\end{document}